\documentstyle[12pt]{article}
\newcommand{\mk}{|{\bf k}|}
\newcommand{\mop}{|{\bf p}|}
\begin{document}

\begin{tabbing}
\`SUNY-NTG-93-21\\
\`May 1993
\end{tabbing}
\vbox to  0.8in{}
\centerline{\Large \bf Heavy dileptons from nonequilibrium QGP}
\vskip 2.5cm
\centerline{\large A.  Makhlin }
\vskip .3cm
\centerline{Department of Physics}
\centerline{State University of New York at
Stony Brook}
\centerline{Stony Brook, New York 11794}
\vskip 0.35in
\centerline{\bf Abstract}
The rate of emission of heavy dileptons from  QGP is  found without
an assumption of its complete thermal equilibrium. We base on the real-time
quantum field kinetic approach [1] and use the expansion up to the second
order with respect to strong coupling constant $g$. The final answer
is not free from the collinear singularities and we show that this is
the actual issue. As a result the main contribution to the rate of the
heavy dileptons production  at $M/T \sim 10$ comes from the process
$q\bar{q}g\rightarrow \gamma^{*}$.

\vskip .25cm
\vfil
\noindent
${^\dagger}$ E-mail address: "makhlin@sbnuc"
\eject

\newpage
\pagestyle{plain}
\addtocounter{page}{-1}

{\bf \Large Introduction}

Transparency of hadronic plasma for electromagnetic signals inspires
hope that they may carry information about the intimate details of interactions
in the quark-gluon media. For this  reason   they are considered as promising
probes of the hot hadronic matter[2]. Initial theoretical efforts were put
basically to the study of an influence of hydrodynamic background on the total
yield of the dileptons [3]. The production rate was taken from the first
Born's  diagram, $q\bar{q}\rightarrow \gamma^{*}$.
The dynamics of quark-gluon interactions was present only virtually
via an assumption that it has led to a thermal distribution of quarks.
 The further improvement of
the microscopic picture also took into account the gluon interactions[4]
up to the $\alpha_s$-order.

The most encouraging result was obtained by Baier, Pire an Schiff [5]. They
found that the rate of emission of the dileptons in the $\alpha_{s}$ order
is defined as well as the first Born's term: all the infrared and mass
singularities has cancelled out in their approach. It was the direct check of
the Lee-Nauenberg theorem for the emission of dileptons from the equilibrium
thermal bath. This result was confirmed in somewhat different manner in [6].

The most important consequence of this study was a statement that even apart
from the factor $\exp(-M/T)$ the rate of dilepton emission is a decreasing
function of $M/T$.

Nevertheless there remain some  questions. First, any deviation
from the detail balance relationships of the true thermal ensemble breaks
a zero balance  of mass singularities. The calculation given below is at
least an example. Second, the rate of dilepton emission in the process
$q\bar{q}g\rightarrow \gamma^{*}$ increases (apart from $\exp(-M/T)$ )
with the growth of the dilepton mass. The effect is due to specific features of
the phase-space volume of the reaction and is not sensitive to
the nature of the
infrared cut-off. The leading role of this
reaction was noticed in [7]. Recently
it was confirmed [8] by numerical calculations. In this paper we give an
analytic estimate.

We confine ourselves to the same first $\alpha_{s}$-order also.
The main difference of this work from the previous ones is that we
consider  the quark gluon plasma without thermal and chemical equilibrium.
We show that  both effects, the absence of collinear singularities
and suppression of the process $q\bar{q}g\rightarrow \gamma^{*}$,
are the artifacts of ideal thermal equilibrium. If the equilibrium  is broken
then at $M/T \sim 10$ the rate of this process overwhelms the Born's rate
while
the rates of all the other processes are suppressed at least by the factor
$T/M$.

An initial goal of this study was to design a tool which could have been used
for calculations of the dilepton emission against the background of partons
distributions generated from cascade simulation of the $A-A$ collision.
Some scenario show [9] that cascade do not reach thermal equilibrium and that
quarks are relatively suppressed with respect to hot glue.   So we used some
flexible analytic approximation for distribution of the partons in favour of
the
opportunity to get an analytic answer and reliable  estimate of the leading
terms.

While choosing this approximation we kept in mind the following scenario
of the heavy ions collision. The initial stage of a collision
at RHIC or LHC energies ($\tau \sim 0.5 fm$) is a region of nucleons
fragmentation and development of the initial parton cascade. Dileptons
are emitted only due to the hard Drell-Yan process. Our region begins a little
bit later when partons are already chaotized and may be described by the
one-particle distributions.

The most general density matrix which simulate any given ahead form
of the one-particle distribution is of the next form,
\begin{equation}
\rho = \prod_{N}\prod_{p,j}e^{-f_{j}(N,p)a^{+}_{j}(N,p) a_{j}(N,p)}
\end{equation}
where $N$ label the space cells on the hypersurface of the initial data
and $n_{j}(N,p)=a^{+}_{j}(N,p) a_{j}(N,p) $ is an operator of the number of
partons of the sort $j$ and quantum numbers $p$ in the $N$-th sell. Thus
we completely neglect all the correlation effects in the phase space of the
partons.

The density matrix (1) allows one to apply the field theory to the
description of the further evolution and results in  a set of habitual
elements like cross-sections, self-energies, vertices. If the functions
$f_{j}(N,p)$ which define the shapes of partons distributions are completely
arbitrary then we are totally confined to  numerical calculations. The
only piece of theoretical study will be connected with the necessity to
transform some singular expressions to the form when numerical calculations
will be unambiguous.

Allover this paper we use the Boltzmann-like distributions
damaged by introduction of specific parameters $\zeta_{Q}$ and $\zeta_{G}$
which
all together will be considered as the measure of nonequilibrium in the
quark-gluon system.We adopt for quarks and gluons
\begin{equation}
n_{F}(p)=\zeta_{Q}e^{-|pu|/T},\;\;\;\;\;\;n_{B}(p)=\zeta_{G}e^{-|pu|/T},
\end{equation}
where $u^\mu$, the 4-velocity of the nonequilibrium partonic media,
fugacities $\zeta$ and temperature are very smooth functions of
space-time coordinates. Scenario of the hot glue [9] leads to
$\zeta_{Q}<\zeta_{G}<1$  and $T$ is rather a formal parameter than a
thermodynamic temperature.  We hope that
distributions (2) are enough flexible to approximate the cascade approaching
to the thermal equilibration.

 An important advantage of this parametrization  is that all the calculations
 can be performed analytically to the very end and result in no more than
 one-dimensional convergent integrals.

 The paper is organized as follows. In Sec.1 we briefly remind the basic
 definitions as they were derived in ref.[1] and trace the way of
 perturbative expansion assuming the weak coupling between the quarks
 and gluons. In Sec.2 we split   the general expression for the
 rate of the dilepton emission  into the pieces which at some heuristic
 level may be called as real processes and radiative corrections.
 In Sec.3  the rate of the real processes is divided into  more
 specified parts, annihilation and Compton and the integration over the
 distributions of initial partons is carried out. We carefully analyze the
  kinematic regions where the further radiative corrections will be needed.
 In  Sec.4 we do the same for the virtual corrections.
 In Sec.5  we perform the remaining integration over the final state of
 the partons. At this step we still have a set of the IR-divergent
 one-dimensional integrals.
 Sec.6 describes the assembling procedure which results in the IR-finite
 answer for the rate of emission. We show that scale which is
 responsible for IR-finiteness of the observable rate is  $T$ (not $gT$!)
 and put forward the arguments why the double logarithms should be
exponentiated
 into the K-factor. In conclusion we analyze the mass singularities and
 show that they are naturally restricted by the amplitude of the
 forward scattering of hard quark in the partonic media. By some chance
 this cut-off coincide with the so-called thermal mass. This mass can not be
 associated with the rest mass of any quasi-particle.

   \renewcommand{\theequation}{1. \arabic{equation}}
\setcounter{equation}{0}
\bigskip
{\bf \Large 1.Perturbative expansion for the rate of emission}
\bigskip

 For sake of completeness  we shall start with reminding the definitions and
 the main framework of calculations[1].
We shall close this section by the formal expansion of the rate of emission
  defined in the scheme of quantum field kinetics up to the first order in
  $\alpha_{s}$.

  The inclusive rate of the dilepton
emission is given by
\begin{equation}
 k_{1}^{0} k_{2}^{0} {{dN_{e^{+}e^{-}}}\over{d{\bf k}_{1}
{d\bf k}_{2}d^{4}x}}=
 -ie_{0}^{2}{{L_{\mu\nu}(k_{1},k_{2})}\over{4(2\pi)^{6}}}
{\bf \Delta}^{\mu\nu}_{10}(-k),
  \end{equation}
where
$k=k_{1}+k_{2}$, $L^{\mu\nu}= k_{1}^{\mu} k_{2}^{\nu}+
k_{2}^{\mu} k_{1}^{\nu}-g^{\mu\nu}(k_{1}k_{2}-m_{e}^{2}) $,
is a trace of lepton spinors,
\begin{equation}
   {\bf \Delta}^{\mu\nu}_{10}(-k)=-i\int d^{4}(x-y)
  \langle T^{+}(A^{\mu}(x)S^{+}) T(A^{\nu}(y)S)\rangle
   e^{-ik(x-y)},
\end{equation}
is a kind of photon Wightman function averaged with
$\rho^{QCD}$ and $e_{0}$ is an electric charge of the lepton.

 We suppose that an explicit separation of
long-range and short-range orders  takes place  and have passed from the
 inclusive cross-sections to the emission rates per unit volume.
That is we have assumed that even not reaching thermal equilibrium
cascade produces at least locally homogeneous distributions of the partons.

    The integral equation  for  ${\bf \Delta}_{10}$  can be effectively
solved in a manner described in [1] and we get the rate of dileptons emission
expressed via polarization operator $\Pi_{10}(-k)$,

\begin{equation}
 k_{1}^{0} k_{2}^{0} {{dN_{e^{+}e^{-}}}\over{d{\bf k}_{1}
{d\bf k}_{2}d^{4}x}}=
 ie_{0}^{2}{{L_{\mu\nu}(k_{1},k_{2})}\over{4(2\pi)^{6}}}
{{\Pi^{\mu\nu}_{10}(-k)} \over {[k^{2}]^{2}} }   ,
\end{equation}
where $[k^{2}]^{-2}$ stands for a product $D_{ret}(k)D_{adv}(k)$
out of the photon mass shell.

  Integrating this distribution over the relative momentum of the leptons we
get
  a distribution of the dileptons over their masses and total momenta:
  \begin{equation}
  {dN_{e^{+}e^{-}}\over d^{4}k d^{4}x}=
  2k_{0}{dN_{e^{+}e^{-}}  \over d^{3} {\bf k} dM^{2}d^{4}x}=
 {i\pi e_{0}^{2}g_{\mu\nu}\Pi^{\mu\nu}_{10}(-k) \over 3(2\pi)^{6}M^{2} }
\end{equation}

   Polarization matrix in these equations is given by an expression
\begin{equation}
  \Pi_{AB}(x,y)=-i(-1)^{A+B} e^{2} \sum_{R,S=0}^{1}(-1)^{R+S}
\! \int \! d \xi d \eta \gamma^{\mu}{\bf G}_{AR}(x,\xi)
 E^{\nu}_{RS,B}(\xi,\eta;y) {\bf G}_{SA}(\eta,x),
\end{equation}
where $e$ stands for the electric charge of  a quark and
a subsequent summation over
the $u$- and $d$-quarks is assumed, $e^{2}=(5/18)e_{0}^{2}$.
It contains an electromagnetic vertex dressed by strong interaction,
\begin{equation}
    E^{\lambda}_{RS,P}(x,y;z)=(-1)^{R+S+P}{
 {\delta [{\bf G}^{-1}(x,y)]_{RS} } \over
  {e \delta {\cal A}_{\lambda}(z_{P}) }  } .
\end{equation}
  The matrix Schwinger-Dyson equation can be solved in
the form,
   \begin{equation}
   [{\bf G}^{-1}]_{AB} = [G_{0}^{-1}]_{AB}-M_{AB},
   \end{equation}
where the matrix of quark self-energy looks as
\begin{eqnarray}
  M_{AB}(x,y)=i(-1)^{A+B}g_{0}^{2}\sum_{R,S=0}^{1} (-1)^{R+S}
   \int d \xi d \eta \times           \\
  \times t^{a} \gamma^{\mu} {\bf G}_{AR}(x,\xi)
   \Gamma^{d,\lambda}_{RB,S}(\xi,y;\eta)
 {\bf D}^{da}_{SA,\lambda\mu}(\eta,x),     \nonumber
\end{eqnarray}
and $\Gamma$ is the vertex of strong interaction.
The inverse Greenian $G_{0}^{-1}$ is nothing but the Dirac operator
 with the "external" field ${\cal A}(x)$,
 \begin{equation}
 [G_{0}^{-1}]_{AB}(x,y)=(-1)^{A}\delta_{AB}\delta(x-y)[i\hat{\partial}_{x}
 +e{\bf t}\hat{\cal A}(x_{A})-m]
 \end{equation}
 Proceeding from the Eq.(1.6 ) and confining ourselves to the two lowest orders
 of the vertex expansion we get
 \begin{eqnarray}
 &&\Pi^{\mu\nu}_{10}(-k)=ie^{2} N_{c} \int{d^{4}p \over (2\pi)^{4} }
 \gamma^{\mu}{\bf G}_{10}(p-k)\gamma^{\nu} {\bf G}_{01}(p)
 +e^{2}g^{2}N_{c}C_{F}\sum_{R,S=0}^{1} (-1)^{R+S} \times \nonumber \\
&&\times \int{d^{4}pd^{4}q \over
 (2\pi)^{8}} D_{SR}(q)\gamma^{\mu}G_{1R}(p-k)\gamma^{\lambda}
 G_{R0}(p+q-k)\gamma^{\nu}G_{0S}(p+q)\gamma_{\lambda}G_{S1}(p)
 \end{eqnarray}
 The second term of this equation is already restricted to the $\alpha_{s}$
order.
 The further expansion of the exact Greenians in the first term is performed
 by means of equations[1],
\begin{equation}
 {\bf G}_{01,10} = G_{01,10}- G_{ret} M_{01,10} {\bf G}_{adv} +
  G_{ret} M_{ret} {\bf G}_{01,10}+ G_{01,10} M_{adv} {\bf G}_{adv} ,
\end{equation}
where  in the order we are interested in all the propagators are to be taken
bare and self-energies should be calculated to the first nonvanishing order in
strong coupling constant. The later are as following,

\begin{equation}
  M_{AB}(s) =i(-1)^{A+B}{{g_{0}^{2}C_{F}}\over {(2\pi)^{4}}}
 \int  d^{4}q\gamma_{\mu}G_{AB}(s+q) \gamma_{\lambda} D^{\lambda\mu}_{BA}(q),
\end{equation}
and
\begin{equation}
  M_{ret,adv}(s) ={{ig_{0}^{2}C_{F}}\over {2(2\pi)^{4}}}
 \int  d^{4}q [\gamma_{\mu}G_{ret,adv}(s+q) \gamma_{\lambda}
D^{\lambda\mu}_{1}(q) +\gamma_{\mu}G_{1}(s+q) \gamma_{\lambda}
D^{\lambda\mu}_{adv,ret}(q)].
\end{equation}

   \renewcommand{\theequation}{2. \arabic{equation}}
\setcounter{equation}{0}
\bigskip
{\bf \Large 2. Separation of real processes and radiative corrections.}
\bigskip

Eqs. (1.11) - (1.14)
 give us the rate of emission in the closed form containing all
contributions of all possible processes. To split them we should notice that an
indication of the type of the process comes from its statistical weight.
Thus we can single out the direct annihilation,
$q\bar{q}\rightarrow\gamma^{*}$, annihilation with the emission or adsorbtion
of a gluon, $q\bar{q}\rightarrow g\gamma^{*}$  and
 $q\bar{q}g\rightarrow\gamma^{*}$, and Compton scattering of quarks on
a gluon, $qg\rightarrow q \gamma^{*}$ and
$\bar{q}g\rightarrow \bar{q}\gamma^{*}$ .

If in the first line of Eq.(1.11)   we substitute   instead of
exact Greenians ${\bf G}_{10,01} $
their unperturbed values we get the first Born's term,  related to direct
process $q\bar{q}\rightarrow\gamma^{*}$:
 \begin{equation}
 \Pi^{\mu\nu}_{Born}(-k)=ie^{2} N_{c} \int{d^{4}p \over (2\pi)^{4} }
 \gamma^{\mu}G_{10}(p-k)\gamma^{\nu} G_{01}(p).
\end{equation}
The diagram of the Born's term is given at Fig1.

 If we insert terms with the off-diagonal elements of the self-energy
 we get two equal terms related to the real processes, which contribute
 to the second Born's approximation term
 \begin{eqnarray}
 \Pi^{\mu\nu}_{a}(-k) &=& 2e^{2}g^{2}N_{c}C_{F}
\int{d^{4}pd^{4}q \over (2\pi)^{8}}
 D_{10}(q)Tr\{\gamma^{\mu}G_{10}(p-k)\gamma^{\nu}   \times  \nonumber \\
 && \times  G_{ret}(p)\gamma^{\lambda}
 G_{01}(p+q)\gamma_{\lambda}G_{adv}(p)\}
 \end{eqnarray}
 The second contribution to the real processes comes from the two equal items
 with $R\neq S$ in the sum in Eq.(1.11).

 \begin{eqnarray}
 \Pi^{\mu\nu}_{b}(-k)&=&-2e^{2}g^{2}N_{c}C_{F}\int{d^{4}pd^{4}q \over
 (2\pi)^{8}} D_{10}(q) \times \nonumber \\
 && \times Tr\{\gamma^{\mu}G_{10}(p-k)\gamma^{\lambda}
 G_{00}(p+q-k)\gamma^{\nu}G_{01}(p+q)\gamma_{\lambda}G_{11}(p) \}
 \end{eqnarray}
In a short while we  shall rearrange $\Pi_{real}=\Pi_{a}+\Pi_{b}$ into
a sum  $\Pi_{real}=\Pi_{em}+\Pi_{abs}+\Pi_{com}$ of the inclusive subprocesses
of the $q\bar{q}$-annihilation with the emission or absorption of the gluons
and the Compton (bremsstrahlung) process. The diagrams of $\Pi_{a}$
and $\Pi_{b}$  are given at Fig.2. The crossed lines present the on-mass-shell
partons with density matrix (1). Their statistical weights originate from the
off-diagonal densities $G_{AB}$ and $D_{AB}$, $A \neq B$, which carry a certain
sign of the energy.

  The  rest items from equations (1.11) and (1.12) are the self-energy and
vertex corrections.

The two remaining terms of the Eq.(1.12), containing $M_{ret}$ and $M_{adv}$
give rise to the radiative corrections of the self-energy type. They
account both for the vacuum effects and for the amplitude of the forward
scattering of quark on the partons produced by the cascade.

 \begin{eqnarray}
 \Pi^{\mu\nu}_{mass}(-k)=e^{2}g^{2}N_{c}C_{F}\int{d^{4}pd^{4}q \over
 (2\pi)^{8}} Tr\{\gamma^{\mu}G_{10}(p-k)
\gamma^{\nu} G_{01}(p) \times \nonumber \\
\times \{ [\gamma^{\lambda}G_{ret}(p+q)\gamma_{\lambda}D_{1}(q)+
 \gamma^{\lambda}G_{1}(p+q)\gamma_{\lambda}D_{adv}(q)]
 G_{ret}(p)+       \nonumber \\
+ [\gamma^{\lambda} G_{adv}(p+q) \gamma_{\lambda} D_{1}(q)+
 \gamma^{\lambda}G_{1}(p+q)\gamma_{\lambda} D_{ret}(q) ] G_{adv}(p)\}
 \end{eqnarray}
 This    expression contains several functions with overlapping singularities
 and it seems better to transform it
 \begin{eqnarray}
 \Pi^{\mu\nu}_{mass}(-k)&=&e^{2}g^{2}N_{c}C_{F}\int {d^{4}pd^{4}q \over
 2(2\pi)^{8}} Tr\{\gamma^{\lambda}G_{s}(p)\gamma^{\mu}G_{10}(p-k)
 \gamma^{\nu} G_{01}(p) \gamma_{\lambda}\} \times \nonumber \\
&& \times (G_{s}(p+q)D_{1}(q)+ G_{1}(p+q)D_{s}(q) )
 \end{eqnarray}

The items with $R=S$ in Eq.(1.11) produce the vertex-type radiative
corrections,
 \begin{eqnarray}
 \Pi^{\mu\nu}_{vert}(-k) &=&-e^{2}g^{2}N_{c}C_{F}
\int{d^{4}pd^{4}q\over (2\pi)^{8}} \times \nonumber \\
 &&\times {\rm Tr}\{\gamma^{\mu}G_{10}(p-k)
\gamma^{\lambda}  G_{00}(p+q-k)\gamma^{\nu}
 G_{00}(p+q)\gamma_{\lambda}G_{01}(p) D_{00}(q)+ \nonumber \\
 &&+\gamma^{\mu}G_{10}(p-k)\gamma^{\lambda}  G_{11}(p+q-k)\gamma^{\nu}
 G_{11}(p+q)\gamma_{\lambda}G_{01}(p) D_{11}(q) \}
\end{eqnarray}
 The vacuum part of the $\pi_{vert}$ is reasonable to calculate just in the
 form given above. It is due to the $T$-independent parts of $G_{00}$
 and $G_{11}$. For the thermal part it is better to transform,
 \begin{eqnarray}
 \Pi^{\mu\nu}_{vert}(-k)=-e^{2}g^{2}N_{c}C_{F}
 \int{d^{4}pd^{4}q\over 4(2\pi)^{8}}
 Tr\{\gamma^{\lambda}G_{01}(p)\gamma^{\mu}
 G_{10}(p-k)\gamma^{\lambda} \times   \nonumber\\
 \times  (G_{s}(p+q-k)\gamma_{\nu}G_{s}(p+q) D_{1}(q)
 +2G_{1}(p+q-k)\gamma^{\nu} G_{s}(p+q) D_{s}(q) )
 \end{eqnarray}
 These transformation present "real" and "virtual" processes in the same
 form: now all thermal diagrams have three cuts!
 The difference is that in radiative corrections
 the extra cut correspond to "elastic scattering on the distribution of the
 partons ".    The typical graphs of this corrections are plotted at
 Fig.3.   Statistical weights of the additional cuts always originate from the
 local density of states $G_{1}$ (or $D_{1}$).They are contributed
both by induced  emission and absorption.

   \renewcommand{\theequation}{3. \arabic{equation}}
\setcounter{equation}{0}
\bigskip
{\bf \Large 3.  Real processes. Integration over the initial states.}
\bigskip

In this section we begin calculation of the real processes.
We do not assume that quarks and gluons
are in thermal equilibrium and use the partons distributions (2) for
the analytic calculations.

At the end of this section the reader will also find physical discussion,
what part of the phase-space is important for the heavy dileptons production.

 In what follows we shall calculate only the distribution of the emitted
dileptons over their total 4-momenta, so we calculate only
$ \pi(k)=g_{\mu\nu}\Pi^{\mu\nu}_{10}(-k)$.

For the Born's term the first integration is the last one,
\begin{equation}
 \pi_{Born}(k)=- {{3ie^{2}}\over{4\pi}}(1+{2m^{2} \over M^{2}} )
 \sqrt{1-{4m^{2} \over M^{2}}} M^{2} e^{-ku/T}.
\end{equation}
In the next perturbative order we have,
 \begin{eqnarray}
\pi_{real}=-{ie^{2}g^{2}N_{c}C_{F} \over  2\pi^{5} } \int d^{4}pd^{4}q
\delta(q^{2})\delta[(p-k)^{2}-m^{2}]\delta[(p+q)^{2}]
   \times \nonumber\\
\times (SW_{em}+SW_{abs}+SW_{com})
 \{1+ {A \over (p^{2}-m^{2})^{2}}+{B \over p^{2}-m^{2}}  +
{ C \over p^{2}-m^{2}+2kq}   \}
\end{eqnarray}
where  we denoted:
\begin{eqnarray}
A=2m^{2}(k^{2}+2m^{2}),\;\;\;\;\;
B &=& 3m^{2}+k^{2}+2(kq)-{ 4m^{4}-(k^{2})^{2} \over 2(kq)} \nonumber \\
C &=& (k^{2}+m^{2})+{ 4m^{4}-(k^{2})^{2} \over 2(kq)}
\end{eqnarray}

The expression in the curly  brackets is nothing but
a sum of the squared moduli of the matrix
elements of the annihilation processes with emission,
$q\bar{q}\rightarrow g\gamma^{*}$, and adsorbtion,
$q\bar{q}g \rightarrow \gamma^{*}$, of a real gluon or the Compton process,
$qg\rightarrow q\gamma^{*}$ and $\bar{q}g\rightarrow \bar{q}\gamma{*}$. Just
for
the purpose of the following calculations it is written not it terms of
habitual
Mandelstam variables $(s,t,u)$. Specification of the process is due to
statistical weights.

The statistical weight of the annihilation process with
emission of a gluon looks as
\begin{eqnarray}
SW_{em}=\theta(k_{0}-p_{0}) \theta(q_{0}+p_{0}) \theta(q_{0})
n_{F}(k_{0}-p_{0})n_{F}(q_{0}+p_{0})[1+n_{B}(q_{0})] \approx \nonumber \\
\approx \theta(k_{0}-p_{0}) \theta(q_{0}+p_{0}) \theta(q_{0})
\zeta_{Q}^{2} e^{-(ku)/T}  (e^{-(qu)/T}+\zeta_{G}e^{-2(qu)/T})
\end{eqnarray}

The statistical weight of the annihilation process with
absorption of a gluon is
\begin{eqnarray}
SW_{abs}=\theta(k_{0}-p_{0}) \theta(q_{0}+p_{0}) \theta(-q_{0})
n_{F}(k_{0}-p_{0})n_{F}(q_{0}+p_{0})n_{B}(-q_{0}) \approx \nonumber \\
\approx \theta(k_{0}-p_{0}) \theta(q_{0}+p_{0}) \theta(-q_{0})
\zeta_{Q}^{2} \zeta_{G} e^{-(ku)/T}.
\end{eqnarray}
For the Compton rate of the dilepton emission the statistical weight equals to
\begin{eqnarray}
SW_{com}=-\theta(p_{0}+q_{0}) \theta(p_{0}-k_{0}) \theta(-q_{0})
n_{F}(p_{0}+q_{0})[1-n_{F}(p_{0}-k_{0})]n_{B}(-q_{0})-   \nonumber \\
-\theta(-p_{0}-q_{0}) \theta(k_{0}-p_{0}) \theta(-q_{0})
n_{F}(k_{0}-p_{0})[1-n_{F}(-p_{0}-q_{0})]n_{B}(-q_{0})- \approx \nonumber \\
\approx  -\zeta_{Q} \zeta_{G} \{ \theta(p_{0}+q_{0}) \theta(p_{0}-k_{0})
\theta(-q_{0}) e^{-pu/T}[1-\zeta_{Q} e^{-(pu-ku)/T}] + \nonumber \\
+\theta(-p_{0}-q_{0}) \theta(k_{0}-p_{0}) \theta(-q_{0})
e^{-ku/T} e^{(qu+pu)/T}[1-\zeta_{Q} e^{(qu+pu)/T}] \}
\end{eqnarray}

The 4-vector $k^{\mu}+q^{\mu}$ is time-like. So we can perform  integration
over $p$ using the Breit reference system where ${\bf k + q}=0$.  The exact
analytic result has the next covariant form:
  \begin{equation}
\pi_{em}={ie^{2}g^{2}N_{c}C_{F} \over  4\pi^{4} } \zeta_{Q}^{2} e^{-ku/T} \int
d^{4}q \delta(q^{2}) \theta(q_{0}) \theta[(k+q)^{2}-4m^{2}] {\cal F}_{a}(kq)
 \{e^{qu/T}+ \zeta_{G} e^{-2qu/T} \} ,
\end{equation}
\begin{equation}
 \pi_{abs}={ie^{2}g^{2}N_{c}C_{F} \over 4\pi^{4}} \zeta_{Q}^{2}\zeta_{G}
e^{-ku/T} \int d^{4}q \delta(q^{2}) \theta(q_{0}) \theta[(k-q)^{2}-4m^{2}]
{\cal F}_{a}(-kq) \end{equation}

 where

\begin{eqnarray}
{\cal F}_{a}(x)=- (1+ { M^{2}+2m^{2} \over x} + {M^{4}-4m^{4}
\over 2x^{2} }) \ln{ 1+ \sqrt{1-4m^{2}/(M^{2}+2x)}
\over  1- \sqrt{1-4m^{2}/(M^{2}+2x)} }+ \nonumber \\
+(1+ { M^{2}+2m^{2} \over x} + {M^{2}(M^{2}+2m^{2})
\over 2x^{2} }) \sqrt{1-{4m^{2} \over M^{2}+2x} }
\end{eqnarray}

We emphasize that  the "nonequilibrium" parameters $\zeta_{Q}$
and $\zeta_{G}$ carry a significant additional information. They allow to
separate contributions of spontaneous and induced processes. This information
would have been  important even in the case of true thermal equilibrium  but it
is completely hidden if we use the detail balance relations at the early stage
of calculations.   For instance, examination of the expression for $\pi_{em}$
immediately shows that the first of the "gluon exponents" originates from quark
and anti-quark distributions and momentum conservation. It describes the
spontaneous emission of a gluon and it is not sensitive to a shape of
 the initial gluons distribution.

Practically, introduction of $\zeta$'s prevents the {\it eventual cancellation}
 of
mass singularities  between real processes and vacuum or thermal mass and
vertex
radiative corrections.  The level of vacuum fluctuations and the rate
of spontaneous processes are fixed {\it de fault} by $\zeta_{vac}=1$.

For the Compton rate we start with a chain of changes of variables aimed to
reduce the statistical weights to the uniform shape. These are
$p\rightarrow -p+k-q$, in the first term and $q \rightarrow -q-p$ in both
terms.
Then we may  perform an exact integration over $p$  using the same Breit
reference system.  It gives

 \begin{equation}
\pi_{compt}=-{ie^{2}g^{2}N_{c}C_{F} \over  4\pi^{4} }
\zeta_{Q} \zeta_{G} e^{-ku/T} \int d^{4}q
\delta(q^{2}-m^{2}) \theta(q_{0})  {\cal F}_{c}(kq)
[e^{-qu/T}-\zeta_{Q}e^{-2qu/T}]
\end{equation}
where $q$ is the momentum of the (anti)quark in the final state and
\begin{eqnarray}
 {\cal F}_{c}(x)= - {4m^{2}+M^{2}-2x-2(M^{4}-4m^{4})/(M^{2}+2x)
  \over 2 \sqrt{x^{2}-M^{2}m^{2}} }
\ln{ m^{2}+x+ \sqrt{x^{2}-M^{2}m^{2}}
\over  m^{2}+x- \sqrt{x^{2}-M^{2}m^{2}}  }  \nonumber \\
+{4(M^{2}+2m^{2}) \over M^{2}+2x }+{(M^{2}+2x)(m^{2}+x) \over
(M^{2}+m^{2}+2x)^{2} } \;\;\;\;\;\;\;\;\;
\end{eqnarray}
Besides an opportunity of analytic integration another advantage of the Breit
system is that it reveals some additional information about the virtual state
of
the quark field. Namely, in annihilation process the energy $p_{0}$ of virtual
state in the Breit system  equals to ratio of squared dilepton mass to twice
energy of initial quarks (or dilepton plus gluon). For the most profitable
configuration of heavy dilepton and soft gluon this is about half  of dilepton
mass; the 4-vector $p_{\mu}$ is space-like and lies very close to the light
cone. For the same reasons the energies of the initial quarks are almost the
same but their 4-momenta are slightly time-like.  For the Compton
process we get the same estimate.  In the most probable case of
heavy dilepton with low momentum ${\bf k}$ the Breit system
 almost coincide with the rest
frame of the media.  So when we will need further mass corrections to the quark
propagators they should be calculated just at these momenta, $p_{0} \sim M/2$
and $p^{2}\sim 0$.

Another important information is that at any finite variable $q$ the very
integrands in Eqs. (3.7,8,10) are divergent at $m \rightarrow 0$. This
mean that the collinear divergencies have explicitly different origin
than the IR-divergencies and should be treated separately.

   \renewcommand{\theequation}{4. \arabic{equation}}
\setcounter{equation}{0}
\bigskip
{\bf \Large 4.  Radiative corrections. Integration over the initial states.}
\bigskip

This section is  technical. It could have been safely berried under the common
"As it can be shown..." if there were no difficulties associated with the
extremely singular form of the integrand.  Being worked out naively they
can essentially change the final answer.

Following Eqs.(2.5) and (2.7) we calculate contributions of quarks and gluons
from the  initial partons distributions to the vertex and self-energy
corrections separately.

 The vertex radiative correction is divided into the two parts contributed
 by the Fermi and Bose parts of the partons distribution,
 $\Pi_{vert}=\Pi_{vert}^{B} + \Pi_{vert}^{F}$:
 \begin{eqnarray}
\pi_{vert}^{B}={ie^{2}g^{2}N_{c}C_{F} \over  4\pi^{5} } \int d^{4}pd^{4}q
\delta(q^{2})\delta[(p-k)^{2}-m^{2}]\delta[(p^{2}-m^{2})]
SW_{vert}^{B}  \times \nonumber\\
 \times {4m^{4}-k^{4}-2kq(m^{2}+k^{2})-4(kp)^{2}+4pq(k^{2}+m^{2}+kq)
\over ((p+q-k)^{2}-m^{2})((p+q)^{2}-m^{2}) }   .
 \end{eqnarray}
 \begin{eqnarray}
\pi_{vert}^{F}={ie^{2}g^{2}N_{c}C_{F} \over  2\pi^{5} } \int d^{4}pd^{4}q
\delta(q^{2}-m^{2})\delta[(p-k)^{2}-m^{2}]\delta[(p^{2}-m^{2})]
SW_{vert}^{F}  \times \nonumber\\
 \times {-m^{2}k^{2}-2kq m^{2}-4(kp)^{2}+2pq(k^{2}+4m^{2}+2kq)
\over (q+k)^{2}-m^{2})(p-q-k)^{2} }   .
 \end{eqnarray}

  In the same way we single out the two parts of the mass corrections,
 $\Pi_{mass}=\Pi_{mass}^{B} + \Pi_{mass}^{F}$:
 \begin{eqnarray}
\pi_{mass}^{B}={ie^{2}g^{2}N_{c}C_{F} \over  2\pi^{5} } \int d^{4}pd^{4}q
\delta(q^{2})\delta[(p-k)^{2}-m^{2}]\delta[(p^{2}-m^{2})]
SW_{mass}^{B}  \times \nonumber\\
 \times {\cal TR}{{\cal P} \over (p^{2}-m^{2}) }
 {{\cal P} \over ((p+q)^{2}-m^{2}) }   .
\end{eqnarray}
 \begin{eqnarray}
\pi_{mass}^{F}={ie^{2}g^{2}N_{c}C_{F} \over  2\pi^{5} } \int d^{4}pd^{4}q
\delta(q^{2})\delta[(p-k)^{2}-m^{2}]\delta[(p^{2}-m^{2})]
SW_{mass}^{F}  \times \nonumber\\
 \times {\cal TR}{{\cal P} \over (p^{2}-m^{2}) }
 {{\cal P} \over (p-q)^{2} }   .
\end{eqnarray}
where ${\cal P}$ means the principal value of the integrals and ${\cal TR}$
 stands for the trace  of spinors from the corresponding Greenians,
 \begin{equation}
 {\cal TR}=2[(p^{2}-m^{2})^{2} +(kq+pq-kp-m^{2})(p^{2}-m^{2})-
 2m^{2}(pq-kp)-2(kp)(pq)+2m^{4}]
\end{equation}

   The corresponding statistical weights for these graphs are
\begin{eqnarray}
 SW_{vert, mass}^{B}=\theta(k_{0}-p_{0}) \theta(p_{0})
n_{F}(k_{0}-p_{0})n_{F}(p_{0})[1+2n_{B}(|q_{0}|)] \approx \nonumber \\
\approx \theta(k_{0}-p_{0}) \theta(p_{0}) \zeta_{Q}^{2}  e^{-(ku)/T}
[1+2 \zeta_{G} e^{-|qu|/T}]
\end{eqnarray}

\begin{eqnarray}
SW_{vert,mass}^{F}=\theta(k_{0}-p_{0}) \theta(p_{0})
n_{F}(k_{0}-p_{0})n_{F}(p_{0})[1-2n_{F}(|q_{0}|)] \approx \nonumber \\
\approx \theta(k_{0}-p_{0}) \theta(p_{0})
\zeta_{Q}^{2}  e^{-(ku)/T}[1 - 2 \zeta_{Q} e^{-|qu|/T}] .
\end{eqnarray}
 One may notice that two of three statistical factors are modified by the
 theta-functions. These are due to the real partons in the initial or final
 states. The third factor works at both signs of energy which means a
 simultaneous account for both absorption of quanta from some state in the
 partonic bath and backward emission to the same state.

The first integration $d^{4}p$ over the initial partons distributions  is
naturally performed in the reference frame of the dilepton. For the vertex
corrections they result in

\begin{eqnarray}
\pi_{vert}^{B}={ie^{2}g^{2}N_{c}C_{F} \over  4\pi^{4} } \zeta_{Q}^{2}
\theta(k^{2}-4m^{2}) e^{-ku/T} \int d^{4}q \delta(q^{2})
 \{ {1 \over 2}+ \zeta_{G} e^{-qu/T} \} \times    \nonumber   \\
 \times [{k^{4}-4m^{4} \over 2(kq)^{2}}\ln{1+v \over 1-v} -v]
\end{eqnarray}
\begin{eqnarray}
\pi_{vert}^{F}={ie^{2}g^{2}N_{c}C_{F} \over  4\pi^{4} } \zeta_{Q}^{2}
\theta(k^{2}-4m^{2}) e^{-ku/T} \times \nonumber \\
\times \int d^{4}q \delta(q^{2}-m^{2})
 \{ {1 \over 2}- \zeta_{Q} e^{-qu/T} \}[{kq-k^{2}-2m^{2} \over
 k^{2}+2kq} v +    \\
 +{(k^{2}+2m^{2})(kq+m^{2})-m^{2}(k^{2}+2kq)/2 \over
 (k^{2}+2kq)\sqrt{(kq)^{2}+k^{2}m^{2}}}
 \ln{   2m^{2}+kq + v\sqrt{(kq)^{2}+k^{2}m^{2}}
 \over  2m^{2}+kq - v\sqrt{(kq)^{2}+k^{2}m^{2}} }].  \nonumber
 \end{eqnarray}

When calculating the self-energy corrections we meet several ill defined
products of the singular functions  like  $x \delta(x){\cal P}(1/x)$.
They need much care and the way to handle them
is described in the Appendix A.
This is hardly the most cumbersome part of calculations. It results in

\begin{eqnarray}
\pi_{mass}^{B}={ie^{2}g^{2}N_{c}C_{F} \over  4\pi^{4} } \zeta_{Q}^{2}
\theta(k^{2}-4m^{2}) e^{-ku/T} \int d^{4}q \delta(q^{2})
 \{ {1 \over 2}+ \zeta_{G} e^{-qu/T} \} \times    \nonumber   \\
 \times \{[1+{m^{2}(k^{2}-2m^{2}) \over k^{2}(kq)}]\ln{1+v \over 1-v}
 +[{2m^{2} \over k^{2}}-{ k^{2}+2m^{2} \over kq} -
 {k^{2}(k^{2}+2m^{2}) \over 2(kq)^{2} }]v \}
\end{eqnarray}
\begin{eqnarray}
 \pi_{mass}^{F}&=& {ie^{2}g^{2}N_{c}C_{F} \over  4\pi^{4} } \zeta_{Q}^{2}
\theta(k^{2}-4m^{2}) e^{-ku/T} \int d^{4}q \delta(q^{2}-m^{2})
 \{ {1 \over 2}- \zeta_{Q} e^{-qu/T} \} \times    \nonumber   \\
 && \times [-{k^{2}+m^{2}+kq \over \sqrt{(kq)^{2}+k^{2}m^{2}}}
 \ln{   2m^{2}+kq + v\sqrt{(kq)^{2}+k^{2}m^{2}}
 \over  2m^{2}+kq - v\sqrt{(kq)^{2}+k^{2}m^{2}} }
 -{2m^{2} \over k^{2}} v ] ,
 \end{eqnarray}
 where we denoted, $v= \sqrt{1-4m^{2}/k^{2}}$.

 Terms corresponding to 1/2 in curly brackets originate from vacuum
 fluctuations. They are both IR- and UV-divergent. Removal of this divergence
is a subject for renormalization. We adopt a standard procedure of the
on-mass-shell renormalization in the asymptotic states both for quark
self-energy and electromagnetic vertex [10]. Thus we insist that vacuum part
of the quark self-energy $M(p)$ has a zero of the second order at
$p^{2}=m^{2}$.
In the same way we adopt that when all three momenta of the vertex function,
\begin{equation}
 \Gamma^{\mu}(k^{2})=\gamma^{\mu}F_{1}(k^{2})+{ \sigma^{\mu\nu}k_{\nu}
 \over 2m }F_{2}(k^{2}),
\end{equation}
are on the asymptotic mass shells of the free
particles then the form-factor $F_{1}(0)=0$.

 This kind of renormalization immediately wash out vacuum part of the quark
self-energy and leaves  the following contribution of quark
electromagnetic form-factors to the dilepton emission as the residue [10]:
\begin{eqnarray}
\pi_{vert}^{vac}= & {ie^{2}g^{2}N_{c}C_{F} \over  4\pi^{3} } \zeta_{Q}^{2}
\theta(k^{2}-4m^{2}) e^{-ku/T} M^{2} \times \;\;\;\;\;\\
\times \{-h(v) \int_{0}^{m}{dq \over q} + & (1+{2m^{2} \over M^{2}})v
-[{3 \over 4}+{m^{2}\over M^{2}}v]\ln{1+v \over 1-v}+
2(1-{4m^{2}\over M^{2}})\int_{0}^{{\rm atanh} v}
{xdx \over \tanh x }\}  \nonumber
\end{eqnarray}
The IR-divergent integral $dq/q$ originates from the soft space-like gluon
exchange between the fermion legs of the electromagnetic vertex and the
function $h(v)$ is given by
\begin{equation}
h(v)=-(1-{4m^{4} \over M^{4}})\ln{1+v \over 1-v} +(1+{2m^{2}\over M^{2}})v
\end{equation}

 An asymptotic of the last integral in (4.14) at $m/M<<1$ looks as
 \begin{equation}
 2\int_{0}^{atanh v} {xdx \over \tanh x } \approx {\pi^2 \over 6}
 +{1 \over 4} \ln^{2}{1+v \over 1-v}
   \approx {1\over 4}\ln^{2}{M^2 \over m^2 }
 \end{equation}
 and produce large negative contribution to the total rate of emission.
 This double logarithm appeared from the collinear configuration of the
 vertex and comes to be infinite for massless quarks. It may be quite
 safe if it is properly balanced with the terms of the opposite sign
 produced by the other processes which are coherent with this one.
 As it was shown in [5,6] such a tiny balance takes place in the ideal
 thermal ensemble. For our distributions it does not happen. But after
 examination of the scales of different processes we shall see that
 this double logarithm should be exponentiated into the K-factor.

   \renewcommand{\theequation}{5. \arabic{equation}}
\setcounter{equation}{0}
\bigskip
{\bf \Large   5.  Final integration.}
\bigskip

There are two time-like 4-vectors in the theory, the 4-momentum of the
dilepton,
$k^{\mu}$, and the 4-velocity of the media, $u^{\mu}$. So we may continue
calculations either in the rest frame of the media or in the rest frame of the
dilepton. We will confine ourselves to the rest frame of the media,
  ${\bf u}=0$, basically because  it is more natural
 to work with the isotropic distribution.

  Let  us consider the  term  responsible for the $q\bar{q}$-annihilation
accompanied by spontaneous emission of the gluon. Its "weight" is
$\zeta_{Q}^{2}\zeta_{vac}= \zeta_{Q}^{2}$. After using of the delta-function
and
one trivial asimutal integration we come to the two-dimensional integral,

\begin{equation}
-\pi T \int_{0}^{\infty} d[e^{ -q/T} ] \int^{1}_{-1} dz q
{\cal F}_{a}(k_{0}q-\mk q z).
\end{equation}
Here and further on  $k_{0}$ stands for $ku$ and $\mk^{2}=(ku)^{2}-M^{2}$.  The
trick which allows one to reduce this double integral to  a simple quadrature
is
as follows [11,12]:

i)change the angular variable $z=\cos\theta$ for the new one,
 $q(k_{0}-\mk z)=y$;

ii) integration with respect to $q$ by parts: in a miraculous way the
integral $dy$ disappears ;

iii) changes  of variables: $q=k_{+}y$ (or $q=k_{-}y$) where
$k_{\pm}=k_{0} \pm \mk$.

The result reads as
\begin{equation}
{\pi T \over \mk} \{ \lim_{\lambda\rightarrow 0}
  \int_{\lambda k_{-}}^{\lambda k_{+}}(1-e^{-y/k_{-}T}) {\cal F}_{a}(y)  dy  +
\int_{\lambda k_{+}}^{\infty}(e^{-y/k_{-}T} - e^{-y/k_{+}T}){\cal F}_{a}(y)
dy\}
 \end{equation}
Though both lower and upper limits of the first integral tend to zero when
$\lambda \rightarrow 0$  the very integral is finite. Indeed, if $f(0)\neq 0$
then
\begin{equation}
\lim_{\lambda \rightarrow 0}
\int_{\lambda k_{-}}^{\lambda k_{+}}{dy \over y}(f(0)+yf'(0)+...)
=f(0)\ln{k_{+} \over k{-}}+ {\cal O}(\lambda),
\end{equation}
and
\begin{equation}
\lim_{\lambda \rightarrow 0}
\lambda \int_{\lambda k_{-}}^{\lambda k_{+}}{dy \over y^{2}}(f(0)+yf'(0)+...)
 ={2\mk \over M^{2}} f(0)+{\cal O}(\lambda)   .
\end{equation}

Foreseeing the future cancellation of soft divergencies we can safely put the
lower limit equal to zero and wright the final result for the annihilation with
the gluon emission in the following form,

\begin{eqnarray}
\pi_{em}={ie^{2}g^{2}N_{c}C_{F} \over  4\pi^{3} } \zeta_{Q}^{2} e^{-ku/T} M^{2}
[(1+\zeta_{G}){k_{+} \over 2\mk} h(v)\ln{k_{+}\over k_{-}} + \nonumber \\
+2 \int_{0}^{\infty} S(y) {\cal F}_{a}(M^{2}y)ydy
+2 \zeta_{G}  \int_{0}^{\infty} S(2y) {\cal F}_{a}(M^{2}y)ydy ].
\end{eqnarray}
The terms with the extra factor $\zeta_{G}$ relate to the induced emission of a
gluon and we denoted,
\begin{equation}
  S(y)=  e^{-k_{0}y/T}{\sinh(\mk y/T) \over (\mk y/T)}
\end{equation}

In a quite similar way we obtain the expression for annihilation with the
absorption of a gluon,
\begin{eqnarray}
\pi_{abs}={ie^{2}g^{2}N_{c}C_{F} \over  4\pi^{3} }
 \zeta_{Q}^{2}\zeta_{G}e^{-ku/T}  M^{2}
 [ (-1+{k_{+} \over 2\mk} \ln{k_{+}\over k_{-}})h(v) + \nonumber \\
 +2 \int_{0}^{(1-4m^{2}/M^{2})/2} {\cal F}_{a}(-M^{2}y)ydy  ]
\end{eqnarray}
where it is helpful to denote
$v(y)=\sqrt{1-4m^{2}/(M^{2}+2y)}$, (now $v(0)=v$)  and rewrite,
\begin{eqnarray}
{\cal F}_{a}(M^{2}y)=- (1+(1+ {2m^{2} \over M^{2}}){1 \over y}
 +(1-{4m^{4} \over M^{4}}){1 \over 2y^{2} }) \ln{ 1+ v(y)
 \over 1- v(y)}  \nonumber \\
 +(1+ (1+{2m^{2}\over M^{2}}){1 \over y} +
(1+{2m^{2} \over M^{2}}){1 \over 2y^{2} })v(y)
\end{eqnarray}

Assembling all gluon contributions to  self-energy and
vertex radiative corrections together we get
\begin{eqnarray}
\pi_{rad}^{B}={ie^{2}g^{2}N_{c}C_{F} \over  4\pi^{3} } \zeta_{Q}^{2}\zeta_{G}
e^{-ku/T} M^{2}[ -{k_{+} \over \mk}h(v) \ln{k_{+}\over k_{-}} + \nonumber\\
+ 2 \int_{0}^{\infty} ({\cal R}(y)+{\cal R}(-y)) S(y)  ydy  ]
\end{eqnarray}
where function ${\cal R}(y)$ in the integrand has a form
\begin{equation}
{\cal R}(y)=\ln {1+v \over 1-v}-(1-{2m^{2} \over M^{2}}v-{h(v) \over 2y^{2}} )
\end{equation}
Equations (5.5), (5.7), (2.9) and the vacuum contribution to the vertex (4.13)
form  a closed system. Though each of them contains IR-divergent
integrals over gluon momentum, when taken together, they produce finite result.

Further calculation of Compton process and radiative corrections from fermion
part of the initial state background meet no infrared problems  at low
momenta of gluons.  So we can safely
choose the simplest way of calculations using the rest frame of a dilepton.
 \begin{equation}
\pi_{compt}=-{ie^{2}g^{2}N_{c}C_{F} \over  4\pi^{3} }
\zeta_{Q} \zeta_{G} e^{-ku/T} 2m^{2} \int_{0}^{\infty}
 {x^{2}dx  \over \sqrt{1+x^{2}}}
{\cal F}_{c}(x)[C(x,T)-\zeta_{Q}C(x,{T \over 2})]
\end{equation}
where
\begin{equation}
  C(x,T)=  e^{{-k_{0}m \over MT}x_{0}  }
  {\sinh({\mk m \over MT}x)   \over ({\mk m \over MT}x)}
\end{equation}
and
\begin{eqnarray}
 {\cal F}_{c}(x)=
{4(M^{2}+2m^{2}) \over M^{2}+2Mmx_{0} }+
{(M^{2}+2Mmx_{0})(m^{2}+Mmx_{0}) \over
(M^{2}+m^{2}+2Mmx_{0})^{2} }-    \\
 - {1 \over  2Mmx}
 \{4m^{2}+M^{2}-2Mmx_{0} - {2(M^{4}-4m^{4}) \over
  M^{2}+2Mmx_{0}} \} \ln{ m^{2}+Mm(x_{0}+x)
  \over  m^{2}+Mm(x_{0}-x)      }  \nonumber
\end{eqnarray}
where $x_{0}=\sqrt{1+x^{2}}$.
The contribution of the Fermi part of intial partons to radiative corrections
is of the next form:
 \begin{equation}
\pi_{rad}^{F}=-{ie^{2}g^{2}N_{c}C_{F} \over  4\pi^{3} }
\zeta_{Q}^{3} e^{-ku/T} 2mM \int_{0}^{\infty} {x^{2}dx  \over \sqrt{1+x^{2}}}
({\cal F}_{f}(x,x_{0})-{\cal F}_{f}(x,-x_{0}))C(x,T)
\end{equation}
where
\begin{eqnarray}
 {\cal F}_{f}(x,x_{0}) = - v ( { { 2 m^{2}+M^{2}-2Mm x_{0}}
 \over { M^{2}+2Mm x_{0} } } -2 )+   \\
+{1 \over x} ( { (M^{2}+2m^{2})(m^{2}+mMx_{0}) \over (M^{2}+2Mmx_{0}) }
-M^{2}-{3 \over 2} m^{2} -mM x_{0}) \ln{ 2m^{2}+Mm(x_{0}+x)
  \over  2m^{2}+Mm(x_{0}-x) }      \nonumber
\end{eqnarray}

   \renewcommand{\theequation}{6. \arabic{equation}}
\setcounter{equation}{0}
\bigskip
{\bf \Large   6. Assembling and cancellation of the IR-singularities.}
\bigskip

 The final answer for the rate of emission is given now as a sum of many terms,
\begin{eqnarray}
   {{dN_{e^+ e^-}}\over{d^{4}kd^{4}x}} =
   {{i\pi e_{0}^{2}}\over{3(2\pi)^{6}M^{2}}} \pi(k),   \\
   \pi(k)=\pi_{Born}+\pi_{em}+\pi_{abs}+ \pi_{com}+\pi_{rad}^{B}+
   \pi_{rad}^{F}+ \pi_{vert}^{vac}, \nonumber
\end{eqnarray}
which are given by Eqs. (3.2), (5.7), (5.10), (5.15) and (4.13). Most
of them are the IR-divergent  one-dimensional integrals. Nevertheless
the whole sum is IR-finite. We show now how it happens and find out the
physical scale which is responsible for this phenomenon.

The  cancellation of the infrared divergencies due to the soft gluons
should be looked at separately for terms with the different powers of
fugacities
$\zeta_{G}$ and $\zeta_{Q}$. The greatest contribution to the rate of the
emission of heavy dileptons comes from the annihilation with the absorption of
a
gluon.  The sum of the leading IR-singular terms from $\pi_{abs}$ and
$\pi_{rad}$ results in the finite integral,
\begin{eqnarray}
\pi_{abs} \approx & {ie^{2}g^{2}N_{c}C_{F} \over  2\pi^{3} }
 \zeta_{Q}^{2}\zeta_{G}e^{-ku/T}  M^{2}
 \int_{0}^{q_{max}}{dq \over q} \{-(1-{4m^{2} \over
M^{2}})(L(-q)-e^{-q/T}L(0))+ \nonumber \\
    & +(1+{2m^{2} \over M^{2}})(v(-q)-e^{-q/T}v(0))\}
\end{eqnarray}
 where $q_{max}=(M^{2}-4m^{2})/2M \sim M/2 $, and
\begin{equation}
v(q)=\sqrt{1-{4m^{2}\over M^{2}+2Mq}},
\;\;\;\;\;\;L(q)=\ln{1+v(q) \over 1-v(q)}.
\end{equation}

The integrand of (6.1) is finite at $q=0$ and takes value $(2/T)\ln(M/m)$
at this point. It is almost linear function and turns to zero at the
upper limit $q_{max}$. This means that radiative corrections $\pi_{rad}$
effectively screen the process at gluon momenta $q<T$ (not $gT$!). The very
integral can be roughly estimated at $M>>m$ as
\begin{equation}
            \sim   2\ln{M \over m}\ln{M\over 2T}
\end{equation}
The result of numerical calculation is drawn at Fig.4   by the bold solid
line in comparison with the thin solid line of the Born's term.
 In the numerical calculations of Ref.[8] no radiative
corrections was introduced and the gluon momentum was cut off at $q\sim gT$.
So the order of magnitude of the effect derived in this way may be different in
the parametric scale (obviously there is no difference at $g\sim 1$).
The finite part of the $\pi_{abs}$ is at least $M/T$ times less than the
leading
term and the rest part of the radiative correction is suppressed by the extra
factor  $\exp(-M/T) $. They are plotted by the dashed lines.

The induced emission along with the corresponding radiative correction has
a statistical weight $\zeta_{G}\zeta_{Q}^{2} $.
The IR-finite combination of the leading terms from $\pi_{em}$ and $\pi_{rad}$
reads as
\begin{eqnarray}
\pi_{em}^{ind} \approx {ie^{2}g^{2}N_{c}C_{F} \over  2\pi^{3} }
 \zeta_{Q}^{2}\zeta_{G}e^{-ku/T}  M^{2}
 & \int_{0}^{\infty}{dq \over q} e^{-q/T}
\{-(1-{4m^{2} \over M^{2}})(L(q)e^{-q/T}-L(0))+ \nonumber \\
    &  +(1+{2m^{2} \over M^{2}})(v(q)e^{-q/T}-v(0))\}
\end{eqnarray}

The integrand of (6.5) takes value $(1/T)\ln(M/m)$ at $q=0$ and the thermal
exponent effectively cut off the upper limit at $q\sim T$. So the order of this
integral is about $-\ln(M/m)$  which is also $M/T$ times less than the leading
term and produces negative contribution to the total rate of emission. One
more term related to this process is positive and suppressed by the factor
$\; \exp(-M/T)$. This curves are plotted at the Fig.5 by the thin solid lines.
This contribution to the rate of emission is small. The Compton process and the
radiative corrections due to $\pi^{F}_{rad}$ do not give much also. They
are plotted at Fig.5 by the dashed and dotted lines respectively, apart from
the factors  $\; \exp(-M/T)$ and $\zeta_{Q,G}$

The spontaneous emission of a gluon by its statistical weight
$\zeta_{Q}^{2} $ is going along with the vacuum vertex radiative correction
$\pi^{vac}_{vert}$. The IR-divergent part of the real process and that of
vacuum vertex together give
\begin{eqnarray}
\pi_{em}^{spont} \approx {ie^{2}g^{2}N_{c}C_{F} \over  2\pi^{3} }
 \zeta_{Q}^{2} e^{-ku/T}  M^{2}\int_{0}^{m}{dq \over q}
\{-(1-{4m^{2} \over M^{2}})(L(q)e^{-q/T}-L(0))+ \nonumber \\
+(1+{2m^{2} \over M^{2}}(v(q)e^{-q/T}-v(0))\}
\end{eqnarray}
It is easily estimated as $-(2m/T)\ln(M/m)$. Sign minus means the negative
contribution of this term to the rate of emission. The factor before $log$ has
replaced a standard $\ln(m/\lambda)$ in course of IR-cancellation. This change
is due to the shape of quark distribution and we have no reasons to introduce
an
artificial mass of a photon. But the collinear singularity,  the big negative
$-\log^{2}(M/m)$, originating from the nonsingular integral (4.15) in the Eq.
(4.13) had survived. All these terms are plotted at Fig.6. The bold solid line
presents (6.6). Thin dotted lines are due do the IR-finite terms
of this process.

The bold dashed line relates to the vacuum vertex.
We notice that in Refs.[5,6] just this term has killed main
contribution (6.4) from the process $3\rightarrow1$. This is  a  remarkable
consequence of the detail
balance in the equilibrium heat bath. In our case this can not happen because
these terms have different powers of fugacities.

An appearance of the $log^{2}$ is a known phenomena and need the further
summation of double logarithms. To understand what should be the result of this
summation we remind that spontaneous emission of the gluons and
the leading ladder
diagrams of the vertex work together and result in the exponential K-factor.
Related only to spontaneous processes, these diagrams have the common
weight $\zeta_{Q}^{2}$ .
We have already got a cancellation of the infrared singularities
 between soft real
gluons and that in the vertex. So  we may estimate the total yield of these two
processes as Born's term times K-factor,
\begin{equation}
K\sim \exp(-{\alpha_{s}C_{F} \over 2\pi}\ln^{2}{M^{2} \over m^{2}}),
\end{equation}
that is as strongly suppressed and even vanishing in the limit of
$m=0$. It is given by the thin dashed line at Fig.6 (with $m\sim gT$).
If we had left the logarithmic term (4.15) as it is then  we could get the
negative total rate of emission in  case  of $\zeta_{G} \leq \zeta_{Q}$ .

  A spontaneous emission of soft gluons which naturally
  accompany the Coulomb interaction of the particles with the hard momenta are
  considered as the  common process with the very interaction of the particles.
  All parameters of the partons distribution are "external" with respect to
this
  block and do not influence its internal dynamics in the gluon sector.
  It even does not seems logical to replace the mass $m$
  in the $K$-factor by the thermal-one.
  Though it came from the quark propagator with big
  momentum $p^{0}\sim\mop\sim M/2$ (see Appendix B) the internal length scale
  of the processes convoluted to the $K$-factor
  is $1/T$ which is less than the length scale $1/gT$
  given by the forward scattering  (see below).

  An exponentiating of the double-$logs$  is in line with the accepted above
  scheme of renormalization of the vacuum parts of the self-energy and the
  vertex. Indeed, an assumption that the UV-divergence is removed by
  subtractions on the asymptotic mass shell means that we consider the
  interaction with the vacuum fluctuations as the dominant one and do not allow
  the interaction with the partons to destroy the mass shell. The existence of
  such a dominant part of interaction is a footing of any normal perturbation
  theory for the many-body systems. In fact, this is the main statement which
  allows to start with the initial distribution of the free partons.

\bigskip
   \renewcommand{\theequation}{7. \arabic{equation}}
\setcounter{equation}{0}
\bigskip

{\bf \Large 7. Conclusion and Discussion.}
\bigskip

  This paper  presents an analytic calculations of the rate of emission
  of heavy dileptons from {\it nonequilibrium} quark-gluon plasma adjusting
  the final result to the needs of computer simulation of the parton
  cascade. Calculations are based on the real-time kinetic approach developed
  earlier [1,7,11].

      An explicit  analytic calculation demonstrate that
  mass singularities remain in the final answer for the rate of the
  heavy dileptons production from the quark gluon plasma. We show
  that the previously obtained opposit result [5,6] is an artifact of the
  true  thermal equilibrium in the partons background. One should not expect
  that  in heavy ions collision this kind of equilibrium will be reached.
  This is in agreement with the known fact that at high momentum transfer
  the double logarithms in the physical answer is rather a rule than an
  exception [12].

  The main practical consequence of this calculation is a conclusion that
  the main process  beyond the first Born's one is annihilation of
  quarks with the absorption of a gluon. The kinematic of this process
  allows dilepton to store an additional amount of the internal energy.
  At $M/T\sim 10$ this process gives as much dileptons
  as the direct process. Ratio of its rate to the rate from the Born's term
  grows up almost linearly with the increasing of the dilepton mass,
  \begin{eqnarray}
     {{dN_{e^+ e^-}}\over{d^{4}kd^{4}x}} =
   {{i\pi e_{0}^{2}}\over{3(2\pi)^{6}M^{2}}}(\pi_{Born}(k)+\pi_{abs}(k)), \\
   \pi_{Born}(k)\approx - {{3ie^{2}}\over{4\pi}}
   \zeta_{Q}^{2}M^{2} e^{-ku/T},  \nonumber \\
   \pi_{abs} \approx  -{ie^{2}g^{2}N_{c}C_{F} \over  2\pi^{3} }
 \zeta_{Q}^{2}\zeta_{G}e^{-ku/T}  M^{2}
   \ln{M^2 \over m^2}\ln{M\over 2T}  .          \nonumber
  \end{eqnarray}
  The processes of annihilation with the spontaneous and induced emission
  of a gluon and the Compton process give a vanishing contribution in
  compare with this process.

  The {\it a priori} unexpected result of the above calculations is the
  cancellation of all those singularities in the thermal graphs which
  originate from soft gluons. This seems to be very significant as
  it reveals some hierarchy of scales and is worth of physical discussion.

   We got a new scale: the effective cancellation
  takes place up to gluon momenta $\sim T$ and this
  natural cut-off  is greater than
  an obviously supposed   $gT$. The later is not so surprising mathematically
  as we compare
  contributions of the graphs of the same perturbation order. Remember that the
  gluonic exponent in the finite combinations like
  $\;\;(L(q)e^{-q/T}-L(0))/q\;\;$ have appeared in the Eqs.(6.1-5) only from
the
  specific form of the quark distributions and from the conservation laws and
  without any connection with the distribution of the gluons in the partons
  cascade.  Notice  that if statistical weight in this combination
 was equal to $1$ (which is the case of process in the free space)
  it would not contain the mass singularity also. This would have been
in complete agreement with the KNL theorem which implies an absence
af any statistical weights for the intermediate states. Probably the true
 thermal equilibrium is  a unique case of populated phase space when
 a complete cancellation of the mass singularities takes place.

  The observed cancellation at $q<T$ means that the two processes,
  spontaneous emission of a real  gluon  and an exchange by the space-like
  gluons, are mutually coherent at all  gluon momenta $q<T$. As only a sum
  of these processes is physical it means that gluons with $q<T$ {\it do not
  participate this process at all} : any attempt of the quark to emit the
  long-wave gluon is immediately killed by the rearrangement of the proper
  fields due to static interaction. This is in agreement with the typical size
  $1/T^3$ of the volume per parton which is free from the
  "third bodies".

  The collinear singularities did survive and they do need the additional
  cut-off. The later can be estimated from the physics of the parallel
  geometry of the front collision. In this geometry the gluon and the massless
  quark interact infinitely long until {\it at least one} interaction with the
  third parton will interrupt this process and destroy the mutual coherence
  of the real and virtual diagrams.

  The external cut-off should come from the smallest of the three lengths.
  The first one is the Compton wave-length $l_{c}\sim 1/m$. The second one is
  the mean free path defined via the scattering cross-section,
  $l_{mfp}\sim 1/g^2 T$. The third one is connected with
  the amplitude of the forward scattering, $l_{fs}\sim 1/gT \sim 1/m_{therm}$.
  As long as we keep quarks massless and consider coupling as the small
  parameter we must chose the length $l_{fs}$ as the smallest one.    So
  whenever we meet the collinear singularity the momenta of both real and
  virtual quarks should be cut off from the below at $\sim gT$.

More formal arguments in favor of the scale $gT$ come from calculation of the
quark self-energy given in Appendix B. We could not use the previous results
by Klimov and Weldon [13] as we deal with essentially different statistical
background. It sufficiently changes (simplifies) an analytic structure of the
quark self-energy. As it is clearly seen from equations (B.6), (B.7) and (B.11)
at high quark momentum the main deviation from the dispersion law of a free
quark is formally simulated by the "thermal mass" of a quark,
$m_{therm}\sim gT$. This does not mean that quark really acquire an additional
rest mass. Indeed, the $m_{therm}^{2}$ dominates in the dispersion equation
(B.6) only when $\mop >>T$ when its influence onto balance of the energy
and momentum is negligible.  We conclude that in the processes with hard
quarks a thermal mass establishes the physical boundary of the possible
coherence length in different processes. This is the main reason why we
prefer to use a cumbersome "length associated with the amplitude of the forward
scattering of the quark in partonic media" instead of a short and simple
"thermal mass".

An actual need of such a boundary appears only in collinear configuration
of the hard processes when its geometric origin is quite evident.

In the kinematic region of the heavy dileptons production (as well as
for hard photons and low-mass dileptons of high energy) an
approximation of the  quark
dispersion law  by formal introduction of thermal mass occurs to be remarkably
exact. The next terms are suppressed by a small factor $\sim T/\mop\sim T/E$.
But this is not true for the case of the small time-like momentum of a quark
when these next terms are very large and even divergent.

A significant piece of work was done by R. Pisarski [14] in attempt to overcome
specific problems of the soft region, $\mop <gT$, by means of selective
resummation of the perturbation series.   This extremely difficult
study is confined to the ideal thermal equilibrium.
Probably a final solution will
demand a nonperturbative approach.

  For the heavy dileptons production when the dilepton mass exceeds any of
  other scales the particular choice
  of the cut-off scale is not so significant within the
  reasonable accuracy. The cut-off mass appears only under the logarithm
  and its variation is not so valuable.

  A choice of the cut-off may cause the
  qualitative changes in calculations of the emission of the low-mass dileptons
  when we are close to the threshold of the existence of the very Born's term
  along with the cascade part of the radiative corrections.  They all appear
  only at $M^{2}>4m^{2}$ ($\sim 4g^2 T^2$ ??) and are absent at the  lower
  masses of the dilepton.

\vskip 1.5cm
I am indebted to G. Brown, E. Shuryak and the Nuclear Theory group at
SUNY at Stony Brook for continuous support.

  I am grateful to  E. Shuryak and I.Zahed for
many fruitful and helpful discussions.
\vskip 1.5cm

   \renewcommand{\theequation}{A. \arabic{equation}}
\setcounter{equation}{0}
\bigskip
{\bf \Large   Appendix A. Singular integrals}
\bigskip

In course of calculations of terms associated with thermal parts of the quark
self-energy we had met extremely singular integrals like
\begin{equation}
I=\int d^{4}p\delta(p^{2}-m^{2}) \delta((p-k)^{2}-m^{2})
\theta(p_{0}) \theta(k_{0}-p_{0}) {{\cal P}\over  p^{2}-m^{2} }F(p)
\end{equation}
were $F(p)$ is a regular function. We choose reference frame $\mk=0$
and get
\begin{equation}
I=\int_{0}^{k_{0}}dp_{0}\int d^{3}{\bf p}\delta(p_{0}^{2}-\omega^{2})
 \delta((p_{0}-k_{0})^{2}-\omega^{2})
 {{\cal P}\over  p_{0}^{2}-\omega^{2} }F(p_{0},{\bf p})
\end{equation}
where $\omega^{2}={\bf p}^{2}+m^{2}$. Using the known formulae,
\begin{eqnarray}
2{{\cal P} \over x}={1 \over x+i\epsilon}+{1 \over x-i\epsilon},\;\;\;\;\;
2\pi i \delta(x)={1 \over x-i\epsilon}+{1 \over x+i\epsilon}\nonumber \\
4\pi i \delta(x){{\cal P} \over x}
={1 \over (x-i\epsilon)^{2}}+{1 \over (x+i\epsilon)^{2} }
\end{eqnarray}
and absorbing all the regular part of the integrand into the function
\begin{equation}
G(p_{0},\omega) ={ F(p_{0},{\bf p}) \over
           (p_{0}+\omega)^{2}(p_{0}-k_{0}-\omega)}
\end{equation}
we get
\begin{equation}
I=-\int {d^{3}{\bf p} \over 8\pi^{2}}\int_{(C)}
{G(p_{0},{\bf p}) dp_{0} \over (p_{0}-\omega)^{2}(p_{0}-k_{0}-\omega)^{2}}
\end{equation}
contour $(C)$ is given by
$(\omega +;(k_0 -\omega)-; \omega -;(k_0 -\omega)+)$.
Now it is easy to see that  the
internal contour integral  differs from zero only at a single point
$\omega -k_{0}=k_{0}$:
 \begin{equation} I=-{1 \over 4} \int d^{3}{\bf p}
G'(\omega,{\bf p}) \delta (\omega-{k_{0} \over 2} )
\end{equation}
 This equation allows to perform the remaining integration over
$d^{3}{\bf p}$ and being applied to  Eqs.(4.3) and (4.4)
  results in Eqs.(4.10) and (4.11)
for  $\pi_{mass}$.

   \renewcommand{\theequation}{B. \arabic{equation}}
\setcounter{equation}{0}
\bigskip
{\bf \Large   Appendix B. Quark self-energy for a nonequilibrium partons
distribution.}
\bigskip

 The invariant decomposition of the "thermal" part of the
 self-energy  of the massive quark
against the background of the initial partons distribution can
be easily found,
\begin{equation}
M_{ret}^{\beta}=-{{\alpha^{s}C_{F}}\over {4\pi \mop^{2}}}
\{4m\lambda + \hat{p}[\eta - (p_{0}^{2}+\mop^{2})\lambda -2p_{0}\xi]
-\hat{u}[p_{0} \eta - p_{0} p^{2}\lambda -2p^{2}\xi] \},
\end{equation}
Where $p_{0}=pu$ and $\mop^{2}=p_{0}^{2}-(pu)^{2}$. The extra $\mop^{-2}$
in quark self-energy  originates from the existence of the preferred reference
frame and provides additional  parameter like $T/\mop$ which is small
in the region actual for heavy dileptons.
In the rest frame of the media we will wright it in the form
\begin{equation}
M_{ret}^{\beta}=W+\gamma^{0}U+ {\bf \gamma p}V
\end{equation}

\begin{eqnarray}
 W={\alpha^{s}C_{F} \over 4\pi \mop^{2} }4m(\lambda_{B}+\lambda_{F}), \;\;\;\;
  U=-2{ \alpha^{s}C_{F} \over 4\pi \mop^{2} }(\xi_{B}+\xi_{F}+
  p_{0} \lambda_{B}), \;\;\;\;\;\;\;\; \\
V= {\alpha_{s} C_{F} \over 4\pi }
     (-{\eta_{B}+\eta_{F} \over \mop^{2}} +
{ p_{0}^{2}+\mop^{2}-m^{2}\over \mop^{2}} \lambda_{B}-
{ p_{0}^{2}-\mop^{2}+m^{2}\over \mop^{2}} \lambda_{F}+
{2p_{0} \over \mop^{2}} (\xi_{B}+\xi_{F}))   \nonumber
\end{eqnarray}
with the following notations,
 \begin{eqnarray}
   \eta_{B}={{ 2}\over{\pi}}\int d^{4}q \delta (q^{2})
n_{B}(q_{0}),\;\;\;\;\;\;
   \eta_{F}={{ 2}\over{\pi}}\int d^{4}q \delta
 (q^{2}-m^{2})n_{F}(q_{0}),\;\;\;\;\;\;\;   \\
     \lambda_{B}={{ 2}\over{\pi}}\int d^{4}q
 {{\delta (q^{2})}\over{p^{2}-m^{2}+2qp}}n_{B}(q_{0}),\;\;\;
 \lambda_{F}={{-2}\over{\pi}}\int d^{4}q
 {{\delta (q^{2}-m^{2})}\over{p^{2}+m^{2}- 2qp}}n_{F}(q_{0}),   \nonumber   \\
\xi_{B}={{2}\over{\pi}}\int d^{4}q
{{\delta (q^{2})}\over{p^{2}-m^{2}+ 2qp}}q_{0}n_{B}(q_{0}),\;\;\;
\xi_{F}={{-2}\over{\pi}}\int d^{4}q
{{\delta (q^{2}-m^{2})}\over{p^{2}+m^{2}- 2qp}}q_{0}n_{F}(q_{0})     \nonumber
\end{eqnarray}

The dispersion equation, ${\rm det}[\hat{p}-m-M_{ret}]=0$, can
be rewritten in the next form,
\begin{equation}
p_{0}^{2}-\mop^{2}-m^{2}+2(p_{0}U+\mop^{2}V-mW)+(U^{2}-\mop^{2}V^{2}-W^{2})=0.
\end{equation}
 The last term of this equation is of the order $\alpha_{s}^{2}$. So if
 we are looking for the roots near their unperturbed positions, relaying
 upon the small value of $\alpha_{s}$, we have every reason to omit it.
 Up to the $\alpha_{s}$-order this equation is as
\begin{equation}
p_{0}^{2}-\mop^{2}-m^{2} - {2\alpha_{s}\over 3\pi}[\eta_{B}+\eta_{F}-
(p^{2}-3m^{2})(\lambda_{B}+\lambda_{F})]=0.
\end{equation}
 In the case when the Lagrangian mass
 obey the inequality,  $m<<|p_{-}|<<T<<M$, the explicit values of the
invariants
 are as follows ($E_{1}$ is the integral exponent),

 \begin{eqnarray}
 \eta_{B}=8\zeta_{G}T^{2},\;\;\;\; \eta_{F}=8\zeta_{Q}T^{2}    \\
 \lambda_{B}= \zeta_{G}L(p) \;\;\;\lambda_{F}= \zeta_{Q}L(p),    \\
 L(p)={T \over \mop}[e^{b_{-}}E_{1}(b_{-})-
 e^{-b_{-}}E_{1}(-b_{-}) - e^{b_{+}}E_{1}(b_{+})
 +e^{-b_{+}}E_{1}(-b_{+}) ]  \nonumber   \\
 \xi_{B}= \zeta_{G}K(p),   \;\;\; \xi_{F}= \zeta_{Q}K(p),  \\
 K(p)={T^{2} \over \mop}[e^{b_{-}}E_{1}(b_{-})(1-b_{-})+
 e^{-b_{-}}E_{1}(-b_{-})(1+b_{-}) - \nonumber \\
 -e^{b_{+}}E_{1}(b_{+})(1-b_{+})-
 e^{-b_{+}}E_{1}(-b_{+})(1+b_{+}) ]    \nonumber
  \end{eqnarray}
  where  $b_{\pm}=(p_{0}\pm \mop )/2T$  and $E_{1}(-x) =E_{1}(e^{-i\pi}x) $
 We are basically interested in the values of quark self-energy at  $M>>T$,
 $p_{+}\sim M$ and $p_{-}\sim 0$. In this region we can use an approximation,

 \begin{equation}
 b_{+}e^{b_{+}}E_{1}(b_{+}) \approx 1 \;\;{\rm and}\;\;\;
  E_{1}(b_{-}) \approx -C_{E}-\ln(b_{-}).
\end{equation}
It gives simple expressions,
\begin{equation}
L(p)\approx -{4T^{2}\over \mop^{2}}-i\pi{2T\over \mop},\;\;\;\;
K(p)\approx  {T^{2}\over \mop}[2-2C_{E}+i\pi -2\ln{p_{-}\over 2T}]
\end{equation}
So, as long as we are interested in corrections to the propagator specific
for the heavy dileptons emission, when $\mop \sim M/2$, the term
$\sim  \alpha_{s}^{2}$  in the Eq.(B.5) is suppressed by the additional powers
of $T/M$.  Up to the next orders in a scale of $T/M$ only the
invariants $\eta$ remain
significant and we can replace Lagrangian mass by the thermal one,
\begin{equation}
m_{therm}^{2}={4g^{2} \over 3\pi^{2}}(\zeta_{Q}+\zeta_{G})T^{2}
\end{equation}

  Contrary to a simple calculation of the Green function for the field in
  QGP the problem of the field quantization in terms of quasiparticles may have
  no solution. Indeed, to be successive on this way we should expand the
  propagator using the {\it complete set of the eigenfunctions of the
  wave equation modified by the self-energy and obeying a certain dispersion
  law}. Even in classical plasma such a set does not exist.  The Green function
  of the Landau kinetic equation allows to solve any problem with the given
  initial data but   a complete set of the solutions of kinetic equation for
  longitudinal waves in a plasma form the so-called Van-Kampen modes which do
  not obey any dispersion law [15].

 \bigskip
 \centerline{\bf REFERENCES}
 \bigskip
  1. A.Makhlin: Preprint  SUNY-NTG-90-11, April 1992.\\
  2.       E. Feinberg: Nuovo. Cimento, 34A (1976) 39;
   E. Shuryak: Sov. J. Nucl. Phys. 28 (1978) 408.              \\
  3. L. McLerran, T. Toimela: Phys.Rev. D31 (1986) 545.       \\
  4. K. Kajantie: P.V. Ruuskanen, Z.Phys. C9 (1981) 341;
     M. Newbert: Z.Phys. C42 (1989) 231.       \\
  5. R.Baier, B. Pire, D. Schiff: Phys.Rev D38 (1988)2819.        \\
  6. T. Altherr, P. Aurenche, T. Becherrawy: Nucl.Phys. B315 (1989)436;
      T. Altherr, P. Aurenche: Z.Phys. C-Particles and Fields 42(1989)99.\\
  7. A.N. Makhlin: Preprint ITP-89-72-E, Kiev, 1989. \\
  8. P. Lichard: Preprint TPI-MINN-92/51-T, October 1992. \\
  9. E.Shuryak :  Phys. Rev. Lett. 68 (1992) 3270.          \\
  10. V.B. Berestetsky, E.M.Lifsits, L.P. Pitaevsky, Relativistic Field
           Theory           \\
  11. A.N. Makhlin: JETP 49(1989)238   \\
  12. Yu.L. Dokshitzer, D.I. Dyakonov, S.I. Troyan: Phys. Reports
58(1980)269.\\
  13. H.A. Weldon: Phys.Rev. D26(1982)1394; Phys.Rev.
           D26(1982)2789; Phys.Rev. D40(1989)2410;
           V.V. Klimov: Yad.Phys. 33(1981)1734  \\
  14. E. Braaten, R.D. Pisarski, T.C. Yuan: Phys. Rev.Lett. 64(1990)2242. \\
  15. N.G. Van Kampen, Physica (Utrecht) 21(1955)49; 23(1957)647; G.
        Ecker, Theory of fully ionized plasmas, Academic Press,
       NY\&London, 1972

\end{document}